\documentclass[aps,prl,twocolumn,superscriptaddress]{revtex4}
\usepackage{mathrsfs}
\usepackage{amssymb, amsbsy, amsmath, latexsym, array, layout,
graphicx,mathrsfs,color,amsfonts,ulem,bm}

\def\be{\begin{equation}}
\def\ee{\end{equation}}
\def\bea{\begin{eqnarray}}
\def\eea{\end{eqnarray}}

\begin{document}

\title{Bloch bound state of spin-orbit-coupled fermions in an optical lattice}

\author{Baihua Gong}
\affiliation{Department of Applied Physics, School of Science, Xi'an Jiaotong University,
Xi'an 710049, Shaanxi, China}
\author{Shuai Li}
\affiliation{Department of Applied Physics, School of Science, Xi'an Jiaotong University,
Xi'an 710049, Shaanxi, China}
\affiliation{Shaanxi Province Key Laboratory of Quantum Information and Quantum
Optoelectronic Devices, Xi'an Jiaotong University, Xi'an 710049, Shaanxi,
China}
\author{Xin-Hui Zhang}
\affiliation{School of Science, Xi'an University of Architecture and Technology, Xi'an 710055, China}
\author{Bo Liu
\footnote {liubophy@gmail.com}}
\affiliation{Department of Applied Physics, School of Science, Xi'an
Jiaotong University, Xi'an 710049, Shaanxi, China}
\affiliation{Shaanxi Province Key Laboratory of Quantum Information
and Quantum Optoelectronic Devices, Xi'an Jiaotong University, Xi'an
710049, Shaanxi, China}
\author{Wei Yi
\footnote {wyiz@ustc.edu.cn}}
\affiliation{Key Laboratory of Quantum Information, University of Science and Technology of China,
Chinese Academy of Sciences, Hefei, Anhui, 230026, China}
\affiliation{Synergetic Innovation Center of Quantum Information and Quantum Physics,
University of Science and Technology of China, Hefei, Anhui 230026, China}

\begin{abstract}
Understanding fundamentals of few-body physics provides an
interesting bottom-up approach for the clarification of many-body properties.
The remarkable experimental progress in realizing spin-orbit coupling
(SOC) in optical Raman lattices offers a renewed thrust towards discovering
{novel few-body features} induced by the interplay between
SOC and optical lattices. Using the Wilson renormalization method to account for high-band effects,
we study the low-energy two-body scattering processes of spin-$1/2$ fermions
in spin-orbit coupled optical lattices.
We demonstrate that, under weak SOC, adding a small lattice potential would destabilize
shallow two-body bound states, contrary to conventional wisdom. On the other hand, when lattice is sufficiently deep, two-body bound states are always stabilized by increasing the lattice depth.
This intriguing non-monotonic behavior of the bound-state stability derives from the competition between SOC and optical lattices, and can be explained by analyzing the low-energy density of states. We also discuss the impact of high-band effects on such a behavior, as well as potential experimental detections.
\end{abstract}

\maketitle


Recent experimental realization of synthetic spin-orbit coupling (SOC) in cold atomic gases
provides unprecedented opportunities for the study of topological matter~\cite{RN44,RN45,RN43,RN52,RN34,RN38,RN47}.
With highly-tunable parameters, cold atomic gases not only serve as perfect simulators for topological phases such as
topological superconductors or Weyl semimetals, but also offer opportunities for the creation of novel topological matter with no counterparts in solids~\cite{RN39,RN40,RN60,RN53,Yi2015}.
An exciting experimental progress of late is the realization of synthetic SOC in Raman lattices~\cite{RN56}. The combination of lattice potential and SOC greatly extends the tunability of cold atoms, which paves the way for more exotic topological phenomena and has therefore stimulated extensive theoretical and experimental studies~\cite{RN35,RN41,RN48}. A fundamental issue here is the characterization of two-body scattering and other few-body processes of the Bloch states, which would shed light on the understanding and control of many-body dynamics for atoms in spin-orbit-coupled lattices. Whereas few-body physics in spin-orbit coupled optical lattices has rarely been discussed before, in light of recent studies where high-band effects can have significant impact on single-particle and many-body properties under large enough SOC~\cite{cui2015,weiPhysRevA_2016}, a theoretical framework capable of depicting few-boy processes beyond the single-band approximation~\cite{PhysRevA_02,PhysRevA_01} is in order.

Here we adopt the Wilson renormalization scheme to systematically introduce high-band effects for atoms loaded into a spin-orbit-coupled lattice potential. Using the renormalized single-band Hamiltonian, we investigate the interplay of SOC, lattice potential and high-band effects on the Bloch bound state, i.e., two-body bound states composed of atoms in the lowest Bloch band. In deep lattices, two-body scattering within the lowest band is dominant when the strength of SOC is small compared to the band gap. However, in shallow lattices, high-band effects can be strong even under weak SOC, and we find that the contribution of virtual-scattering processes into higher bands give rise to dramatic changes in the binding energy of bound states. We further identify a parameter regime where increasing the lattice depth would destabilize bound states, contrary to the conventional wisdom that adding SOC or lattice potential to attractively interacting fermionic atoms would enhance the formation of bound states. Such an anomaly is due to the competition between SOC and lattice potential, and can be understood by analyzing the low-energy density of states (LDOS) of the system. We then discuss the impact of high-band effects on such abnormal phenomena. Our results systematically reveal the interplay of SOC, lattice potential and high-band effects in two-body processes of fermions in spin-orbit coupled optical lattices, which would form building blocks for future explorations of exotic many-body phenomena, such as the BEC-BCS
crossover~\cite{RN51,RN55}, in these systems.

\textit{Effective model $\raisebox{0.01mm}{---}$}
We consider an attractively interacting two-component ultracold Fermi gas loaded in a cubic optical lattice in the presence of a Weyl-type SOC. The single-particle physics can
be described by the following Hamiltonian
\begin{equation}
{\mathbf{h}}(\mathbf{r}) = \frac{\mathbf{p}^2}{2m} + V_{OL}(\mathbf{r}) + \lambda
{\mathbf{p}} \cdot {\boldsymbol{\sigma}},  \label{single_H}
\end{equation}
where the lattice potential $V_{\mathrm{OL}}({\mathbf{r}})=-V_0
[\cos^2(k_{L}x)+\cos^2(k_{L}y)+\cos^2(k_{L}z)]$, with {$V_0$ the} lattice depth
and $k_{L}$ the wavevector of the laser fields. The corresponding
lattice constant is defined as $a_L=\pi/k_{L}$. $\lambda {\mathbf{p}} \cdot {%
\boldsymbol{\sigma}}$ is the three-dimensional isotropic Weyl-type SOC
and ${\sigma}_{x,y,z}$ are the Pauli matrices. In the following, we
shall use the recoil energy $E_R=\hbar^2 \pi^2/2m a_L^2$ as the energy
unit, and the lattice depth is characterized by a dimensionless
quantity $v=V_0/E_R$.

\begin{figure}[t]
\begin{center}
\includegraphics[width=8 cm]{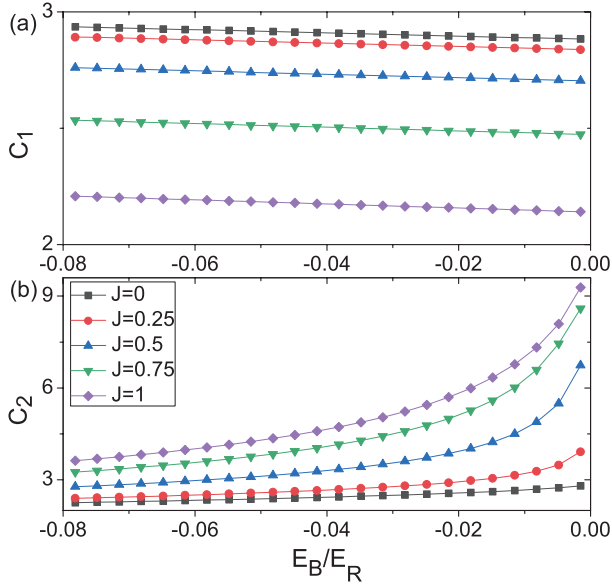}
\end{center}
\caption{ Contributions to the Bloch bound state
arising from higher bands (a) and from the
lowest band (b), respectively, for various SOC strengths $J$
defined as $J\equiv 2m a_L \protect\lambda /\hbar^2 \protect\pi$.
The lattice depth here is chosen as $v=2$.}
\label{c1_c2}
\end{figure}

In order to analyze two-body problems in
spin-orbit coupled optical lattices, it is natural to start with the case
without SOC, in the basis of Bloch states $|n, \mathbf{k}, s \rangle$, where $n$,
$\mathbf{k}$, and $s$ label the band indices, Bloch wave vector within the first
Brillouin zone and spin, respectively. SOC couples the Bloch states according to
\begin{equation}
\langle n, \mathbf{k}, s| \lambda {\mathbf{p}} \cdot {\boldsymbol{\sigma}}
|n^{\prime }, \mathbf{k}^{\prime }, s^{\prime }\rangle = \lambda
\delta_{nn^{\prime }} \delta_{\mathbf{k} \newline
\mathbf{k}^{\prime }} \sum_{\alpha=x,y,z} \overline{k}_{n,\alpha}
\sigma_{\alpha}^{ss^{\prime }}  \label{Matric_SOC}
\end{equation}
with $\overline{k}_{n, \alpha} = \langle n, \boldsymbol{k},s | \left(-i\hbar\frac{\partial}{\partial \boldsymbol{r}_{\alpha}}\right) |
n, \boldsymbol{k}, s \rangle$ ($\alpha = x, y, z$). We focus on the case
with weak SOC compared to the band gap of optical lattices without SOC, in contrast to previous studies where the focus has been in
regimes with large SOC~\cite{cui2015,weiPhysRevA_2016}.
Under weak SOC, we neglect interband matrix elements in Eq.~\eqref{single_H}, while keeping intraband couplings.
The single-particle dispersion of Eq.~\eqref{single_H} is then
\begin{equation}
\epsilon_{n\mathbf{k} \pm } = \varepsilon_{n\mathbf{k}} \pm \lambda
\overline{k}_n \equiv \varepsilon_{n\mathbf{k}} \pm \lambda ( \overline{k}%
_{n,x}^2 + \overline{k}_{n,y}^2 + \overline{k}_{n,z}^2 )^{1/2}
\label{Single_dispersion}
\end{equation}

To investigate two-body scatterings for fermions in such a spin-orbit coupled optical lattice, we start from the interacting Hamiltonian
\begin{eqnarray}
{\mathbf H}&=& {\mathbf{H}}_0+{\mathbf{H}}_{\text{int}}  \notag \\
&=& \mathbf{h}_1(\mathbf{r}_1) + \mathbf{h}_2(\mathbf{r}_2) + U(\mathbf{r}_1 -%
\mathbf{r}_2), \label{twobody_H}
\end{eqnarray}
where $\mathbf{h}_1$, $\mathbf{h}_2$ represent single-particle Hamiltonians of the colliding atoms
as shown in Eq. ~\eqref{single_H}. The short-range interaction between fermions of different spin species is captured by $U(\mathbf{r})=U_0 \delta^3(\mathbf{r})$. Here $U_0$ is the bare
interaction strength associated with the effective low-energy scattering
length $a_s$ via {the standard renormalization condition} $1/U_0  = m/4 \pi a_s- \int d\mathbf{k} m/(2\pi)^3k^2$.

We now evaluate the interaction matrix elements between different Bloch states.
First, we introduce the basis $\{ |m, \mathbf{k} \uparrow; n, -\mathbf{k} \uparrow
\rangle, | m, \mathbf{k} \uparrow; n, -\mathbf{k} \downarrow \rangle, |m,
\mathbf{k} \downarrow ; n, -\mathbf{k} \uparrow \rangle, |m, \mathbf{k}
\downarrow; n, -\mathbf{k} \downarrow \rangle$, with $|\alpha;
\beta \rangle =  \left(|\alpha \rangle \otimes |\beta
\rangle - |\beta \rangle \otimes |\alpha \rangle \right)/\sqrt{2}$. Here, ${\alpha}$ ($%
{\beta}$) stands for Bloch states in a given spin sector with $k_z>0$.
We then classify interaction matrix elements into the following three different types:
(I) $|0, \mathbf{k}, s; 0, -\mathbf{k},
\overline{s} \rangle \leftrightarrow |0, \mathbf{k}^{\prime }, s^{\prime },
; 0, -\mathbf{k}^{\prime }, \overline{s^{\prime }} \rangle$, (II) $|n,
\mathbf{k}, s; n, -\mathbf{k}, \overline{s} \rangle \leftrightarrow
|n^{\prime }, \mathbf{k}^{\prime }, s^{\prime }; n^{\prime }, -\mathbf{k}%
^{\prime }, \overline{s^{\prime }} \rangle$ with $n\neq 0$ or $n^{\prime
}\neq 0$, (III) $|n, \mathbf{k}, s; n, -\mathbf{k}, \overline{s} \rangle
\leftrightarrow |m^{\prime }, \mathbf{k}^{\prime }, s^{\prime }; n^{\prime
}, -\mathbf{k}^{\prime }, \overline{s^{\prime }} \rangle $ with $m^{\prime
}\neq n^{\prime }$, where $s, \overline{s}$ refer to opposite spins
respectively. The first type describes the scattering within the
lowest band, playing the most dominant role in scattering processes.
The corresponding matrix elements can be approximated as $M
U_0/\Omega$, where $M$ is determined by the overlap of Bloch states and $\Omega$ is
the volume. The second type includes the most important scattering processes
involving higher bands, which plays the sub-dominant role. The matrix
elements can be approximated as $U_0/\Omega$. The third type
describes scattering processes within two different bands and plays a
minor role compared to the first two types.
We have numerically checked that, whereas the first two types of matrix elements are of the same order of magnitude, the last one is typically smaller than the first two types by three orders of magnitude.
To simplify the problem, we thus neglect the third type in the following studies.

The Hamiltonian in Eq. ~\eqref{twobody_H} can thus be expressed as
\begin{widetext}
 \begin{equation}
 {\mathbf H}_0^{n,\boldsymbol{k}}=
 \left(
 \begin{array}{cccc}
 2\varepsilon_{n\boldsymbol{k}} & 0 & 0 & 0 \\
 0 & 2\varepsilon_{n\boldsymbol{k}} & 0 & 0 \\
 0 & 0 & 2\varepsilon_{n\boldsymbol{k}}& 0 \\
 0 & 0 & 0 & 2\varepsilon_{n\boldsymbol{k}} \\
 \end{array}
 \right)
 +
\lambda \left(
\begin{array}{cccc}
0 & - \overline{k}_{n,x} +i  \overline{k}_{n,y}   &  \overline{k}_{n,x} - i  \overline{k}_{n,y}   & 0 \\
- \overline{k}_{n,x} -i  \overline{k}_{n,y}   & 2  \overline{k}_{n,z}  & 0 &  \overline{k}_{n,x} -i  \overline{k}_{n,y}   \\
\overline{k}_{n,x} +  i  \overline{k}_{n,y}   & 0 & -2  \overline{k}_{n,z}  & - \overline{k}_{n,x} +i  \overline{k}_{n,y}  \\
0 &  \overline{k}_{n,x} +i  \overline{k}_{n,y}   &  -\overline{k}_{n,x} - i  \overline{k}_{n,y}   & 0 \\
\end{array}
\right)
\label{H_0}
 \end{equation}
 \end{widetext}
and
\begin{equation}
{}_{m,\boldsymbol{k}}\langle i|U|j\rangle _{n,\boldsymbol{k}^{\prime }}=%
\begin{cases}
(-1)^{\delta _{ij}+1}\frac{U_{0}}{\Omega } &
\begin{array}{l}
(m\neq 0\text{ or }n\neq 0)\  \\
\&\ (i,j=2,3)%
\end{array}
\\
(-1)^{\delta _{ij}+1}M\frac{U_{0}}{\Omega } &
\begin{array}{l}
(m=n=0)\  \\
\&\ (i,j=2,3)%
\end{array}
\\
0 & i=1,4\text{ or }j=1,4%
\end{cases}
\label{U}
\end{equation}
Here, the basis $\{ |i \rangle_{n\boldsymbol{k}} \}$ is defined as $\{ |n, \boldsymbol{k}
\uparrow; n, -\boldsymbol{k} \uparrow \rangle, | n, \boldsymbol{k} \uparrow;
n, -\boldsymbol{k} \downarrow \rangle, |n, \boldsymbol{k} \downarrow ; n, -%
\boldsymbol{k} \uparrow \rangle, |n, \boldsymbol{k} \downarrow; n, -%
\boldsymbol{k} \downarrow \rangle, k_z>0 \}$, which can be written as
$\{ |i \rangle_{n\boldsymbol{k}}, \ i=1,2,3,4\} $ for convenience.

\begin{figure}[t]
\begin{center}
\includegraphics[width=8 cm]{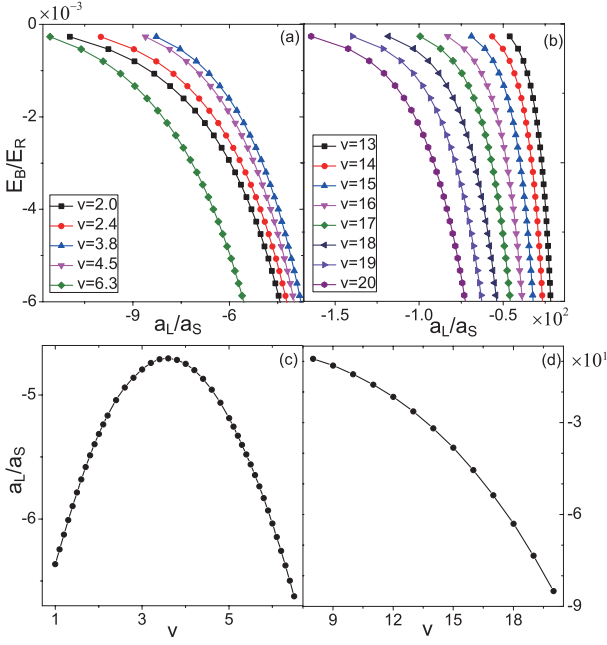}
\end{center}
\caption{(a)(b) Binding energy $E_B$  as a function of the s-wave scattering length
$a_s$ in shallow (a) and deep (b) lattices, respectively. The energy reference is chosen as
twice the minimum of the single-particle energy
$2\min({\epsilon_{0\mathbf{k} - }})$. (c)(d) The scattering length $a_s$ as a function of lattice depth $v$
with a fixed shallow binding energy $E_B =-0.004E_R$ in shallow (c) and
deep (d) lattices, respectively. The threshold lattice depth determined by the non-monotonic variation of ${a_L}/{a_s}$ in (c) is $v=3.8$. Here we choose the strength of SOC $J=0.65$.} \label{soc20_xiao}
\end{figure}

\textit{Effective theory for lowest Bloch band through
Wilson Renormalization$\raisebox{0.01mm}{---}$}
To understand the importance of scattering processes involving higher bands, we
employ the Wilson renormalization method~\cite{RN42} to construct an
effective theory in the lowest Bloch band. The basic idea of such a renormalization procedure is to infer
properties of a system, whose description requires a large number of
basis functions, by progressively reducing the dimension of the Hilbert
space spanned by the preserved basis functions within a certain energy range.
For this purpose, we first separate the
Hilbert space into two subspaces $A$ and $B$, which stand for
the lowest $n=0$ Bloch band and all the other bands with $n \neq 0$, respectively. Then the Hamiltonian ${\mathbf H}$ in
Eq.~\eqref{twobody_H} can be written as follows
\begin{equation}
{\mathbf H} = \left(
\begin{array}{cc}
{\mathbf H}^{AA} & {\mathbf H}^{AB} \\
{\mathbf H}^{BA} & {\mathbf H}^{BB}%
\end{array}
\right) = \left(
\begin{array}{cc}
{\mathbf H}^{AA}_0 + U^{AA} & {\mathbf H}^{AB} \\
{\mathbf H}^{BA} & {\mathbf H}^{BB}_0 + U^{BB}%
\end{array}
\right)
\end{equation}
The renormalized effective Hamiltonian ${\mathbf H}_{\text{eff}}$
for the lowest Bloch band, which is entirely in
subspace $A$, can be obtained through the Wilson renormalization method~\cite{Supply}.

\begin{figure}[t]
\begin{center}
\includegraphics[width=8 cm] {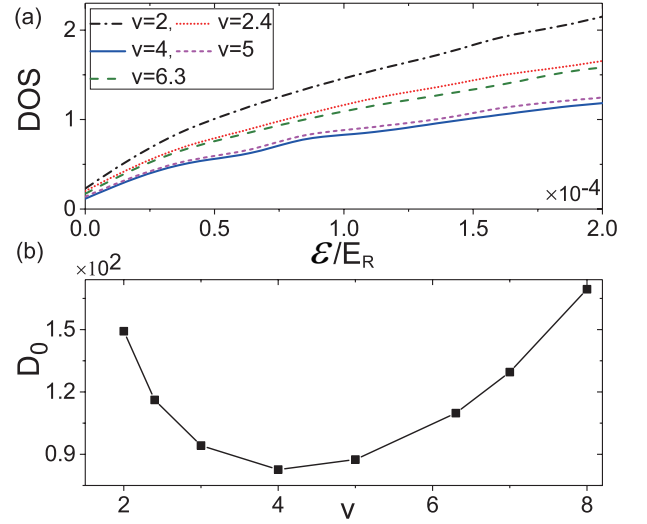}
\end{center}
\caption{(a) LDOS as a function of single-particle
energy $\varepsilon$ for various lattice depths $v$. (b) The non-monotonic
behavior of LDOS driven by an increasing lattice depth. Here the LDOS is modeled by a
square-root type function $D_0\sqrt{\varepsilon}$. The threshold lattice
depth determined by the non-monotonic variation of LDOS
in (b) is $v=4$. The strength of SOC is the same as in Fig.~\ref{soc20_xiao}.}
\label{DOS}
\end{figure}

\textit{Bloch bound state $\raisebox{0.01mm}{---}$}
We are now in position to characterize bound states within the lowest Bloch band and discuss high-band effects.
Starting from the renormalized effective
Hamiltonian ${\mathbf H}_{\text{eff}}$ introduced
above, we define the Green's function as
$G=\left(
E-{\mathbf H}_{\text{eff}}\right)^{-1}$. The binding energy $E_B$ of the
Bloch bound state {is} obtained from  poles of the Green's
function defined above, which satisfies~\cite{Supply}
\begin{align}
    \frac{a_L}{a_s}  = C_1(E_B)  -C_2(E_B)
    \label{bound}
\end{align}
where
\begin{align}
    C_1(E_B)  & =   \frac{4\pi a_L }{m \Omega} \left[ \sum_{  |\boldsymbol{k}| < \Lambda^*}  \frac{1}{  k^2/m } +   \sum_{n>0, \boldsymbol{k}   \in \mathrm{BZ}}      \frac{1}{2}  \left ( \frac{1}{E_B -2\epsilon_{n\boldsymbol{k}+}  } \right. \right. \notag \\
&+ \left. \left. \frac{1}{  E_B -2\epsilon _{n\boldsymbol{k}-} }     \right)   \right] \\
    C_2(E_B) & = \frac{4\pi a_L  M }{\eta m  \Omega}     \sum_{\boldsymbol{k}   \in \mathrm{BZ}} \frac{1}{2}   \left( \frac{1}{2\epsilon_{0\boldsymbol{k},+}  - E_B } + \frac{1}{  2\epsilon _{0\boldsymbol{k},-} -E_B }  \right)
\end{align}
Here, $C_1$ and $C_2$ capture contributions to the Bloch bound state
arising from the higher $n>0$ bands and lowest $n=0$ band
respectively. A high-energy cutoff $\Lambda^*$ is introduced to facilitate numerical calculations, which does not affect final results. Note that the binding energy $E_B$ is defined with
respect to twice the minimum of the single-particle energy $2\min({\epsilon_{0\mathbf{k} - }})$.

In the deep-lattice limit, high-band effects
are negligible, and we
recover equations for bound states in the absence of
SOC by dropping $C_1$ in Eq.~\eqref{bound}~\cite{RN70,RN71}. In contrast, for shallow
lattices, both SOC and lattice potential have dramatic impact on the formation of
Bloch bound state.
As shown in Fig.~\ref{c1_c2}, we
find that in shallow lattices both high-band effects and the
intraband scatterings within the lowest band are important. These
two kinds of contribution are generally comparable and neither of
them can be neglected. This suggests that the simple lowest-band approximation
in previous studies~\cite{PhysRevA_02,PhysRevA_01} need to be improved. Interestingly,
we notice that high-band effects, signaled by $C_1$, become more significant with decreasing SOC.
{This is because the minimum of single-particle energy decreases with increasing SOC, in which case shallow bound states
are less affected by higher bands.}

We numerically solve the binding energy $E_B$ from Eq.~\eqref{bound}.
As shown in Fig.~\ref{soc20_xiao}, we calculate $E_B$ as a function of the $s$-wave scattering length $a_s$ for different lattice
depths. For $a_s<0$, there is always a Bloch bound state regardless of the lattice depth. This is quite different from the case without SOC, where a threshold in the lattice depth exists, below which there are no Bloch bound states~\cite{RN70,RN71}. In deep lattices [Fig.~\ref{soc20_xiao}(b)], $|E_B|$ decreases monotonically with increasing lattice depth at a fixed $s$-wave scattering length. In contrast, for a shallow lattice, there is an abnormal competition between SOC
and lattice potential. As shown in Fig.~\ref{soc20_xiao}(a), for a fixed $a_s$, the binding energy $|E_B|$ first increases and then decreases
with increasing lattice depth. The competition between SOC and lattice potential in the formation of bound state can be seen more clearly in Fig.~\ref{soc20_xiao}
(c), where a threshold lattice depth exists, below which one has to increase the interaction strength to achieve the same binding energy with increasing lattice depth. This indicates destabilization of the bound state by the lattice potential, contrary to the conventional wisdom. Above the threshold, as the lattice depth increases, the same binding energy can be achieved with weaker interaction strength. This is consistent with properties of the bound state in the deep lattice limit, as shown in Fig.
~\ref{soc20_xiao} (b) and (d).

\begin{figure}[t]
\begin{center}
\includegraphics[width=8 cm] {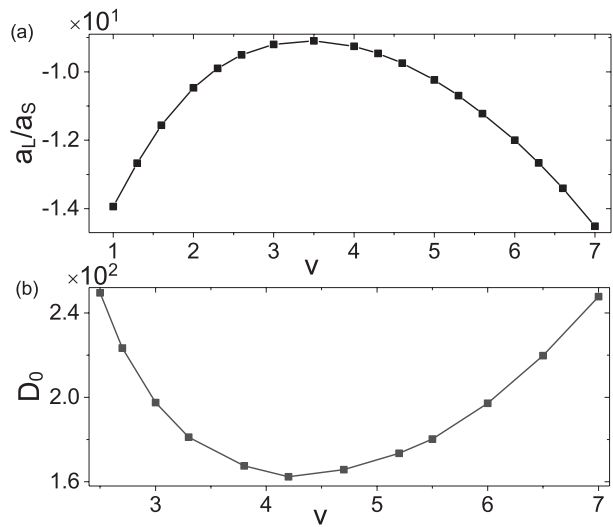}
\end{center}
\caption{(a) The scattering length $a_s$ as a function of lattice depth $v$
with a fixed binding energy $E_B =-0.0001E_R$. (b) The non-monotonic
behavior of LDOS driven by an increasing lattice depth. The threshold lattice
depth determined by the non-monotonic variation of ${a_L}/{a_s}$ in (a) and LDOS
in (b) are $v=3.5$ and $v=4.2$, respectively. Here we fix $J=0.4$.}
\label{soc_04}
\end{figure}

Such an anomalous interplay of SOC and lattice potential
on the formation of the Bloch bound state can be understood by considering
LDOS of a single fermion moving
in a spin-orbit coupled optical lattice.
Heuristically, the symmetric
Weyl-type SOC strongly enhances the LDOS
in a three-dimensional uniform system, with the low-energy scaling
$\sim 1/\sqrt{\varepsilon}$~\cite{RN50},
where $\varepsilon$ is the single-particle dispersion.
Whereas without SOC, the LDOS near the energy minima of the lowest band in a
three-dimensional cubic lattice has the scaling $\sim\sqrt{\varepsilon}$~\cite{RN42}. As the scaling of LDOS is intimately connected with the stabilization of bound states, one may then expect a competition between SOC and lattice potential on the formation of Bloch bound states.

The non-monotonic variation of LDOS with increasing lattice depth is confirmed in Fig.~\ref{DOS}(a).
We then numerically analyze such a non-monotonic behavior in LDOS using a squre-root-type scaling $D_0\sqrt{\varepsilon}$. Importantly, the turning point in the LDOS, as illustrated in Fig.~\ref{DOS}(b), roughly coincides with the threshold lattice depth in Fig.~\ref{soc20_xiao}(c). This confirms that the change in LDOS is the main reason behind the anomalous behavior in the bound-state formation. Whereas the variation of LDOS is determined by properties of the lowest band, {impact of high-band effects are manifested as the difference in lattice-depth thresholds calculated from $a_L/a_s$ and from LDOS alone.}
More explicitly, by comparing Fig.~\ref{DOS}(b) and Fig.~\ref{soc20_xiao}(c), we find that high-band effects reduce the lattice-depth threshold. When decreasing the strength of SOC, high-band effects are strengthened, which is demonstrated by a larger reduction in the lattice-depth threshold shown in Fig.~\ref{soc_04}(a)(b). This suggests that high-band effects weaken the competition between SOC and lattice potential, and
{tend to enlarge the regime where Bloch bound states are stabilized with increasing lattice depth.}

\textit{Discussion and Conclusion $\raisebox{0.01mm}{---}$}
Adopting the Wislon renormalziation scheme, we derive an effective single-band Hamiltonian in the weak-SOC regime, which allows systematic study of few-body physics in spin-orbit coupled lattices where high-band effects can be important. Our construction is quite general and can be systematically extended to study other types of SOC in different dimensions.
For example, in three-dimensional cubic lattices under weak two-dimensional Rashba-type SOC, the non-monotonic behavior of LDOS driven by the lattice depth is also found, indicating a similar competition between SOC and
lattice potential therein. In contrast, when considering weak one-dimensional SOC, the LDOS would be monotonically enhanced when increasing the lattice depth.

Whereas the impact of high-band effects on the single-particle properties have already been observed in recent experiments~\cite{1710.00717,1804.08226} {based on a new realization scheme of SOC~\cite{Xiong_jun_2018}}, the intriguing interplay between SOC, lattice potential and high-band effects can be demonstrated through measuring the binding energy of Bloch bound states. This can be achieved, for example, by measuring the distance between the atomic peak and the onset of the molecular feature in a typical radio-frequency (rf) spectrum~\cite{wuPhysRevLett_2012} or an rf dissociation spectrum~\cite{jinnature_2003}.

\textit{Acknowledgment $\raisebox{0.01mm}{---}$} This work is supported by the Natural Science Foundation of China Grant Nos. (11774282,11522545). W. Y. acknowledges support from the National Key R\&D Program (Grant Nos. 2016YFA0301700, 2017YFA0304800)

\bibliographystyle{apsrev}
\bibliography{wenxian}


\onecolumngrid

\renewcommand{\thesection}{S-\arabic{section}}
\setcounter{section}{0}  
\renewcommand{\theequation}{S\arabic{equation}}
\setcounter{equation}{0}  
\renewcommand{\thefigure}{S\arabic{figure}}
\setcounter{figure}{0}  

\indent

\begin{center}\large
\textbf{Supplementary Material:\\ Bloch bound state of spin-orbit-coupled fermions in an optical lattice}
\end{center}

\section{Details of Renormalization  method}
The renormalized effective Hamiltonian for the lowest Bloch band can be obtained through the following method
\begin{align}
{\mathbf H}_{\text{eff}}(E) &= {\mathbf H}^{AA} + {\mathbf H}^{AB} \frac{1}{E - {\mathbf H}^{BB}} {\mathbf H}^{BA} \notag \\
& = {\mathbf H}^{AA}_0 + U^{AA} + {\mathbf H}^{AB} \frac{1}{E -
{\mathbf H}^{BB}} {\mathbf H}^{BA} \label{Heff2}
\end{align}
The effect of high bands is characterized by the last term in
Eq.~\eqref{Heff2}, where $E$ is the
energy spectrum and $\left(E-{\mathbf H}^{BB}\right)^{-1}$  can
be determined following the procedure detailed below. We first project the
Hamiltonian in Eq.~\eqref{twobody_H} into the subspace $B$ and
obtain
\begin{equation*}
{\mathbf H}^{BB} = {\mathbf H}^{BB}_0 + U^{BB}
\end{equation*}
Then the above equation can be written as
\begin{equation}
E - {\mathbf H}^{BB}_0 = (E - {\mathbf H}^{BB}_0 - U^{BB} ) + U^{BB}
\label{H_BB}
\end{equation}
Multiplying both sides of Eq.~\eqref{H_BB}
by $1/(E -{\mathbf H}^{BB}_0 )$ from the left and $1/(E -{\mathbf
H}^{BB}_0 - U^{BB})$ from the right, the Eq.~\eqref{H_BB} can be
rewritten as
\begin{equation}
\frac{1}{ E - {\mathbf H}^{BB} } = \frac{1}{ E - {\mathbf H}^{BB}_0
} + \frac{1}{ E - {\mathbf H}^{BB}_0 } U^{BB}\frac{1}{ E - {\mathbf
H}^{BB} } \label{GBB}
\end{equation}

Then the last two terms in Eq.~\eqref{Heff2} can be expressed in the the basis
$\{ |i \rangle_{n=0,\boldsymbol{k}} \}$  as follows
\begin{eqnarray}
&&{}_{0, \boldsymbol{k}} \langle i| U^{AA} + {\mathbf H}^{AB}
\frac{1}{E - {\mathbf H}^{BB}}
{\mathbf H}^{BA} | j \rangle_{0, \boldsymbol{k}^{\prime }} \notag \\
&=&
\begin{cases}
(-1)^{\delta_{ij}+1 } \frac{ \tilde{U} }{\Omega} , \ \  & i, j=2,3 \\
0 & i=1,4 \text{ or } j=1,4%
\end{cases}%
\end{eqnarray}

and ${\mathbf H}_0^{AA}$ can be read from ${\mathbf
H}_0^{n=0,\boldsymbol{k}}$ as shown in Eq.~\eqref{H_0}, where
$\tilde{U} $ is defined as
\begin{align}
    \tilde{U} =    \frac{ M U_0 \eta^{-1} }{ 1- \frac{U_0}{\Omega} u }
    \label{U_yuanshi}
\end{align}
with $\eta = \left[ 1 - (1 - \frac{1}{M}) \frac{ U_0}{\Omega} u \right]^{-1}$ and
$u = \sum_{n>0, \boldsymbol{k} \in \mathrm{BZ} } \frac{1}{2} \left( \frac{1}{%
E-2\epsilon_{n\boldsymbol{k}+} } + \frac{1}{ E -2\epsilon _{n\boldsymbol{k}%
-} } \right)$. Here $\mathrm{BZ}$ labels the first Brillouin zone of the
optical lattice.

\section{Binding energy of Bloch bound state}
The binding energy $E_B$ of the Bloch bound state is obtained from
poles of the Green's function defined from the renormalized effective
Hamiltonian ${\mathbf H}_{\text{eff}}$ in
Eq.~\eqref{Heff2}, which yields the following relation
\begin{equation}
  \frac{1}{ \tilde{U} } - \frac{1}{\Omega} \sum_{\boldsymbol{k} \in \mathrm{BZ}} \frac{1}{2}  \left( \frac{1}{E_B-2\epsilon_{0k+}  } + \frac{1}{  E_B -2\epsilon _{0k-} }   \right) = 0
  \label{pole}
\end{equation}
Substituting Eq.~\eqref{U_yuanshi} into  Eq.~\eqref{pole}, we then obtain Eq.~\eqref{bound}.


\end{document}